\documentclass[11pt]{article}
\usepackage{moriond,epsfig}

\def\Journal#1#2#3#4{{#1} {\bf #2}, #3 (#4)}

\def\PRD{{\em Phys. Rev.} D}

\def\be{\begin{equation}}
\def\ee{\end{equation}}
\def\bea{\begin{eqnarray}}
\def\eea{\end{eqnarray}}

\begin{document}
\vspace*{4cm}
\title{LAMBDA-INFLATION VERSUS OBSERVATIONS}

\author{ E.V. MIKHEEVA }

\address{Astro Space Center, P.N.Lebedev Physical Institute, 
84/32 Profsoyuznaya st.,\\
Moscow 117997, Russia}

\maketitle\abstracts{
Lambda-inflation is tested against CMB and LSS observational data.
The constraints for inflationary parameters are considered.}

\section{Introduction}
An increasing precision of measurements in observational cosmology 
requires the adequate accuracy of predictions in the early Universe 
theory. It is particularly true for numerious recent and on-going 
experiments to measure the CMB angular anisotropy 
(BOOMERanG~\cite{boom}, MAXIMA~\cite{maxima}, DASI~\cite{dasi}, 
VSA~\cite{vsa}, CBI~\cite{cbi}, ARCHEOPS, MAP).  

The inflationary paradigm is still the only viable theory of 
the early Universe. There is no doubt that inflation explains 
the observable features of the Universe much better than other 
theories (for example, string or ekpyrotic ones). One of such 
features is well established flattness of 
the Universe~\cite{boom,maxima,dasi,vsa}. Other important predictions
of inflation is related to the production of cosmological perturbations.
If the potential of inflaton is determined, the spectra of scalar 
and tensor perturbations can be derived by means of any of existing 
formalisms. Here I use the historically first 
one proposed in~\cite{qscalar}. 

A majority of papers dedicated to the numerical analysis of 
cosmological models considers scale-free cosmological perturbation 
spectra whereas almost all inflationary models predict more complicated 
forms of them. In addition, in many inflationary models the cosmic 
gravitational waves significantly contribute into the large scale CMB 
anisotropy and have to be taken into account. 

This problem may be solved in two ways. The first is testing given 
inflationary models. This way requires accurate calculations of 
scalar and tensor perturbation spectra. The precision of such method 
is related with the validity of slow-rolling approximation. The second way 
is phenomenological one. It is based on the assumption that a density 
perturbation spectrum can be approximated by power law, $k^n$, 
where $n$ may be constant or slowly depending on scale. In the latter 
case a new parameter, the running $\partial n/\partial\ln k$, appears 
(see, e.g., \cite{hanstead} and references wherein). 

My present analysis is related with the first way.

\section{Lambda-inflation}
I call by Lambda-inflation a model with a single scalar 
field $\varphi$ and the following inflaton potential:
\be
V(\varphi)=V_0+\frac{1}{\kappa}\lambda_\kappa\varphi^\kappa,
\ee
where $V_0$, $\lambda_\kappa$, $\kappa$ are constants (see 
also~\cite{linf,linf2}). I will concentrate on a model with $\kappa=4$.

This model is often referred in literature as ``hybrid 
inflation''~\cite{hybridlinde}. I reserve the latter name for 
the case of Lambda-inflation, where some particular way of 
the decay of the effective Lambda-term ($V_0$) is assumed.

It is convinient to introduce two new parameters. The first of them, 
$c$, is as follows:
\be
c\equiv\frac 12\sqrt{\frac{V_0}{\lambda}}
\ee
(here and below $\lambda\equiv\lambda_\kappa$). $c$ is closely 
related to the critical field value where both terms of the potential 
are equal to each other ($\varphi_{cr}=2\sqrt{c}$). The second parameter, 
$k_{cr}$, is a scale when $\varphi$ reaches $\varphi_{cr}$. 
The slow-roll approximation is valid for $c\gg 1$. In this case
the spectra of perturbations are as follows\footnote{Anywhere 
the index S means ``scalar perturbations'', T -- ``tensor perturbations''.}:
\be
{\rm S}:\;\;\;\;\;
q_k=\frac{H}{2\pi\sqrt{2\gamma}}=\frac{\sqrt{2\lambda/3}}{\pi}
\left(c^2+x^2\right)^{3/4},
\ee
\be
{\rm T}:\;\;\;\;\;
h_k=\frac{H}{\pi\sqrt 2}=\frac{2c\sqrt{\lambda/3}}{\pi}
\left(1+\frac{x}{\sqrt{c^2+x^2}}\right)^{-1/2},
\ee
where $H$ is the Hubble constant, $\gamma\equiv -\dot H/H^2$, 
$x\simeq\ln\left(k/k_{cr}\right)$ (see details in \cite{linf,linf2}).
Figure 1 illustrates these formulas.

\begin{figure}
\centerline{\psfig{figure=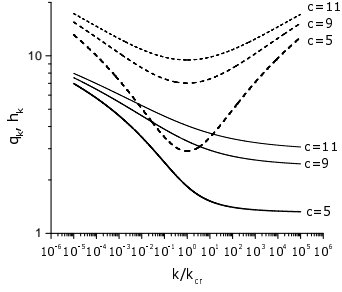,height=1.5in}}
\caption{Scalar ($q_k$, dash lines) and tensor ($h_k$, solid lines) 
perturbation spectra in Lambda-inflation (arbitrary normalization).}
\end{figure}

\section{Cosmological model}
Cosmological parameters can be separated into few groups. One of 
them contains inflationary parameters describing spectra of 
the cosmological perturbations (usually, they are the slopes, 
$n_S$, $n_T$, and amplitudes, $q_k$, $h_k$). Another set of 
parameters describes matter and energy in the Universe
(the matter density parameter, $\Omega_m$, $\Lambda$-term, 
$\Omega_\Lambda$, the baryon abundance, $\Omega_b$, the massive 
neutrino abundance, $\Omega_\nu$, the number of massive neutrino
species $N_\nu$, etc.). Another important cosmological parameter is 
the value of Hubble constant, 
$H_0=100h\;{\rm km}\;{\rm s}^{-1}{\rm Mpc}^{-1}$. 
This short list of cosmological parameters may be expanded.

The complete analysis of complex cosmological model is a very 
hard task: one has to use both a large number of free parameters 
and all available experimental data related to the determination 
of cosmological model ($\Delta T/T$, cluster abundance data, 
$Ly_\alpha$ data, and others).

Let us simplify this task. Firstly, let us fix values of the cosmological 
parameters describing dark matter and energy. Here I assume 
$\Omega_\Lambda=0.65$, $\Omega_c=0.29$, $\Omega_b=0.05$, $\Omega_\nu=0.01$, 
$N_\nu=3$, $h=0.65$\footnote{This cosmological model considered with 
scale-invariant spectrum of density perturbations and negligible value of 
cosmic gravitational waves satisfies both COBE and $\sigma_8$ tests.}. 
Secondly, let us take only two observational tests, namely, COBE 
measurent of $\Delta T/T$ which determines the amplitude of $\Delta T/T$ 
at 10th spherical harmonic ($C_{10}$), and $\sigma_8$ which normalizes
the spectrum of density perturbations.

As for $C_{10}$, we have some gravitational wave contribution coming 
from inflation. It is standardly described by T/S parameter: 
T/S$\equiv C^{\rm T}_{10}/C^{\rm S}_{10}$. The characterictic scale of COBE 
is $k_{COBE}\equiv k_1\simeq 1.5\times 10^{-3} h\; {\rm Mpc}^{-1}$. 
As for $\sigma_8$, the characteristic scale is $k_{\sigma_8}\equiv 
k_2\simeq 1.5\times 10^{-1}h\;{\rm Mpc}^{-1}\simeq 100k_1$. 

After simple calculations we derive the following equation:
\be
\left(1+{\rm T/S}\right)\left(\frac{q_{k_1}}{q_{k_2}}\right)^2=1,
\ee
The numerical solution of this equation can be approximated by linear fit:
\be
\ln\frac{k_1}{k_{cr}} = 0.32\,c-1.92,
\ee
the error is in the last digits of both numbers. An important consequence
of this solution is that the critical scale should be higher than the COBE 
scale, so the former is either close to or behind the present horizon.

\begin{figure}
\centerline{\psfig{figure=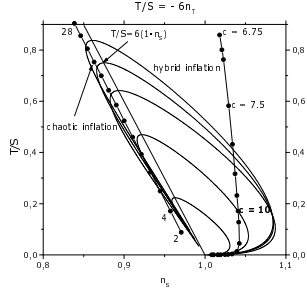,height=1.5in}}
\caption{T/S versus $n_{\rm S}$ and $c$ in Lambda-inflation.}
\end{figure}

Figure 2 presents the corresponding T/S evaluated by means of 
the famous consistency relation T/S$\simeq -6n_{\rm T}$. The inflationary 
models, located along the line with solid circles (marked by "c=..."), 
satisfy both observational tests mentioned above. Assuming that T/S 
is not large (T/S$<0.5$) we may constrain $n_{\rm S}$ ($1<n_{\rm S}<1.05$) 
and inflationary parameter $c$ ($c>8$). 

Recall that $n_{\rm S}$ is estimated at COBE scale
and can change at different scales (for example, $n_{\sigma_8}$).
Figure 3 demonstrates running parameter $\partial n_{\rm S}/\partial\ln k$. 
We see that the density perturbation slope at LSS scale ($k_2$) depends on 
$c$ and can vary from 1 till 1.2.

\begin{figure}
\centerline{\psfig{figure=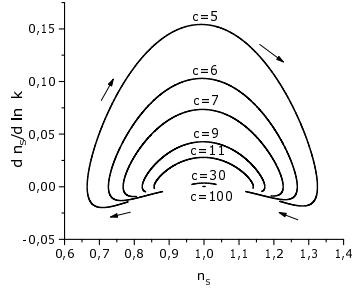,height=1.5in}}
\caption{$\partial n_{\rm S}/\partial\ln k$ versus $n_{\rm S}$ 
in Lambda-inflation.}
\end{figure}

\section{Discussion}
The demonstrated example of $\Lambda$-inflation favors slightly 
blue spectra of density perturbations ($n_{\rm S}{}^>_\sim 1$). 
However, phenomenological constraints for $n_{\rm S}$ based on 
LSS observational data, while converging to $n_S\simeq 1$, are still 
uncertain about the sign of ($n_S-1$). E.g., galaxy cluster data
prefer $n_S>1$ \cite{novos}, whereas galactic surveys indicate
slightly red spectra ($n_{\rm S}<1$) (e.g. \cite{hanstead}). Certainly, 
better data are required to delimit the slope of the fundamental 
spectrum. If future analysis reveals a blue spectrum, 
the $\Lambda$-inflation model will gain strong support as 
the theory of the very early Universe.

\section*{Acknowledgments}
This work was supported by RFBR 01-02-16274 and Russian State contract 
40.022.1.1.1106. I thank the Organizing Committee for hospitality.

\section*{References}

\end{document}